# TABLE OF CONTENTS

# Data Converter Design Space Exploration for IoT Applications: An Overview of Challenges and Future Directions

Buddhi Prakash Sharma, Anu Gupta, and Chandra Shekhar

*Abstract*— Human lives are improving with the widespread use of cutting-edge digital technology like the Internet of Things (IoT). Recently, the pandemic has shown the demand for more digitally advanced IoT-based devices. International Data Corporation (IDC) forecasts that by 2025, there will be approximately 42 billion of these devices in use, capable of producing around 80 ZB (zettabytes) of data. So data acquisition, processing, communication, and visualization are necessary from a functional standpoint. Indicating sensors & data converters are the key components for IoT-based applications. The efficiency of such applications is truly measured in terms of latency, power, and resolution of data converters motivating designers to perform efficiently. Sensors capture and covert physical features from their chosen environment into detectable quantities. Data converter gives meaningful information and connects the real analog world to the digital component of the devices. The received data is interpreted and analyzed with the digital processing circuitry. Ultimately, it is used as information by a network of internet-connected smart devices. Because IoT technologies are adaptable to nearly any technology that may provide its operational activity and environmental conditions. But the challenges occur with power consumption as the complete IoT framework is battery operated and replacing a battery is a daunting task. So the goal of this chapter is to unveil the requirements to design energy-efficient data converters for IoT applications.

# I. Introduction

**Data converters** are a key component of "Internet of Things (IoT) devices". They are commonly utilized to bridge the analog world to the digital parts of gadgets, or vice versa. They are crucial for modern sensors including audio, biomedical, and automotive sensors, which require real-time data to be digitized for filtering, monitoring, signal processing, and analysis, as well as conversion back to the analog domain. These converters are critical for deciphering the detected data. Before the twentieth century, Kevin Ashton's concept of letting computers know everything about "things" was dubbed as IoTs. He planned to utilize computers with sensor technologies and radio frequency identification (RFID) [1] to collect meaningful data and identify an area without the assistance of humans. The term "Internet of Things" [2-3] refers to a network of physical objects-"things"-embedded with sensors, actuators, signal conditioning circuitry, data converters, and signal processing that allow them to remotely communicate and share data with other devices.

IoT is rapidly becoming one of the most exciting technological developments in history [4]. It has been deployed extensively in a scenario such as automotive industries, transportation, smart societies, agriculture, etc. Its uses in daily life, include sensor networks, biological records, and array signal processing [5-6]. The application of wireless sensor networks (WSNs) for health monitoring gives advantages over wire-line systems such as minimizing the risk of infections/failures, increasing mobility, and optimizing the cost and operability. Ultra-low-power (ULP) circuit is critical for extending battery life in portable and self-contained applications. IoT is an opportunity for nations and people to better have control of their data and especially, give value to local information and data. The development of data converters was primarily driven by two factors: technology and applications. The technology utilized by an electronic system is primarily determined by the system's requirements: nanoscale technologies are advantageous to speed and performance. Analog-to-Digital Converters (ADCs) and their counterpart with high precision are required in industrial process control, medical instruments, and data collection systems (10-bit or more). Speed is less important than precision in these applications because the signal bandwidth is often in the tens of kHz range.

In any case, an appropriate source of energy capable of meeting the application's requirements is necessary. Radio has the highest energy needs of any component of an IoT node. Taking care of these requirements can aid in the planning and implementation of the energy harvesting system. [7]. The industrial IoT entry point of interest begins with the edge node of sensing and measuring. This is the node where the real world connects with informatics. Connected factories can detect a wide range of data that will be utilized to make critical decisions. The fundamental aspects of the various stages of IoT architecture can be explained by: sensing, measuring, interpreting, and connecting data as described in fig. 1 [8]. IoT sensors are predominantly analog. IoT edge device design can look to be as difficult as squaring the circle. It has a sensor that connects to the internet. The sensor signal is sent through an amplifier or filtering circuit to an analog signal processing system. The signal is digitized by connecting the output to an A/D converter. The data is analyzed using this signal, which is delivered to the digital processing circuitry. IoT technologies are adaptable to nearly any technology that may provide its operational activity and environmental conditions.

These days, numerous businesses across a variety of industries are utilizing this technology to streamline, enhance, automate, and regulate various processes. We then demonstrate a few of the IoT's unexpectedly useful practical uses. The peculiar industrial application needs (real-time decisions) will speak about the bandwidth and dynamic range of the sensor and actuator that will be required in the analog front end (AFE) of the IoT architecture. The AFE of this chain will be part of the analog domain before sensed data is converted to a digital domain and transmitted signal to the cloud.

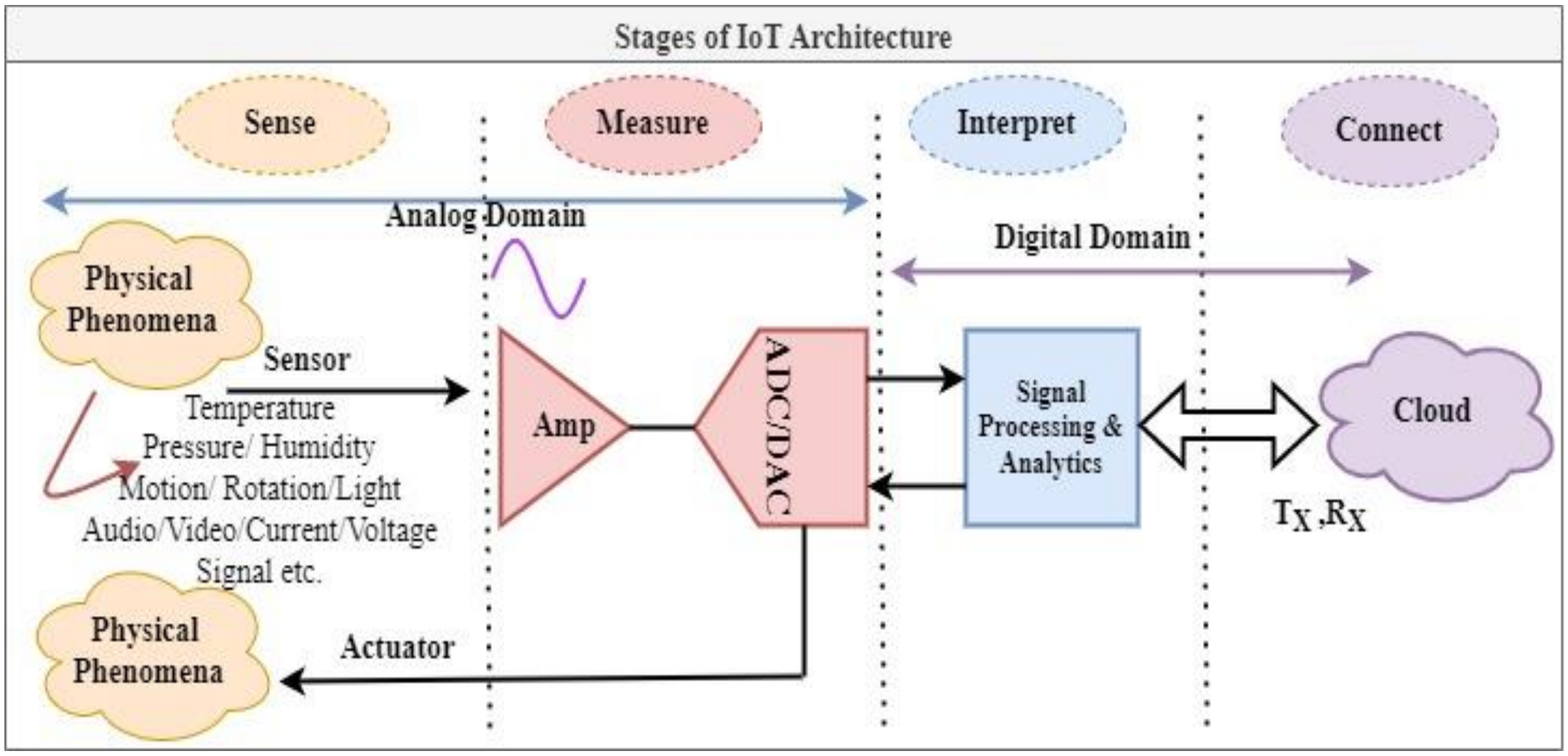

Figure 1: Internet of things (IoT) sensor node

In most cases, signal sampling or mathematical interpretation of signal data is the most energy-intensive task for such a system shown in fig.1 To minimize the energy cost of the device at sensing and measuring nodes, strategies are being developed [9]. Thus, data converters are one of the vital elements in IoT. They are extensively used to interact with the surrounding world with the digital part of the devices. They are crucial for contemporary sensors such as audio, light, motion, biomedical and automotive sensors for which physically collected information is required to be digitized for filtering, monitoring, signal processing, and analysis enhancement. Such kinds of data converters play a critical role in understanding the sensed data.

In this chapter, we demonstrate the role of data converters in IoT, especially in the healthcare sector, and also describe the current challenges. This chapter is structured as follows: Section II illustrates the various data converter architectures and fundamentals. Section III introduces the trends and benchmarking summary. This section illustrates the role of specific data converters in healthcare IoT. It also gives a brief about current challenges and future directions for data converters in IoT.

## II. Data Converter Overview

### *2.1 Converter Architectures*

While there are many different data converter architectures available, the Successive Approximation Register (SAR) ADC, Delta-Sigma ($\Delta$-$\Sigma$) ADC, Flash ADC, and Pipelined ADC are the most popular ones. This section also illustrates various Digital-to-Analog (DAC) topologies.

#### *2.1.1 Successive Approximation Register Analog-to-Digital Converter*

The fundamental blocks of a SAR A/D converter are a track-and-hold (T&H) or sample-and-hold (S&H) circuit, a comparator, digitally controlled logic & registers, and a digital-to-analog block as shown in fig. 2. The first SAR ADC algorithm implementation has been traced back to 1940 at the Bell Labs [10]. McCreary and Gray created the charge redistribution (CR), which is the SAR ADC used today, in 1975 at the University of California, Berkeley [11]. The SAR A/D converter working principle is based on a “binary search algorithm”. The conversion takes place throughout multiple clock cycles. The first step of the conversion is to sample the input voltage. The S&H function is integrated with the DAC in most of the SAR architectures. In the second step, the comparator block compares the sampled value and the output voltage of the DAC to determine the current bit. The conversion starts on each bit from the MSB-LSB by taking one bit at a time. At last, the difference between the input voltage and DAC voltages goes toward zero showing the completion of the conversion process. Therefore, N conversion steps are required for an N-bit

ADC. The internal clock frequency must be at least N-fold than the sampling frequency if the SAR ADC is managed by synchronous logic.When medium-high conversion rates (about hundreds of MS/s) are needed, asynchronous logic is preferred because it does not require the same high-speed clock, hence reducing power consumption[12].

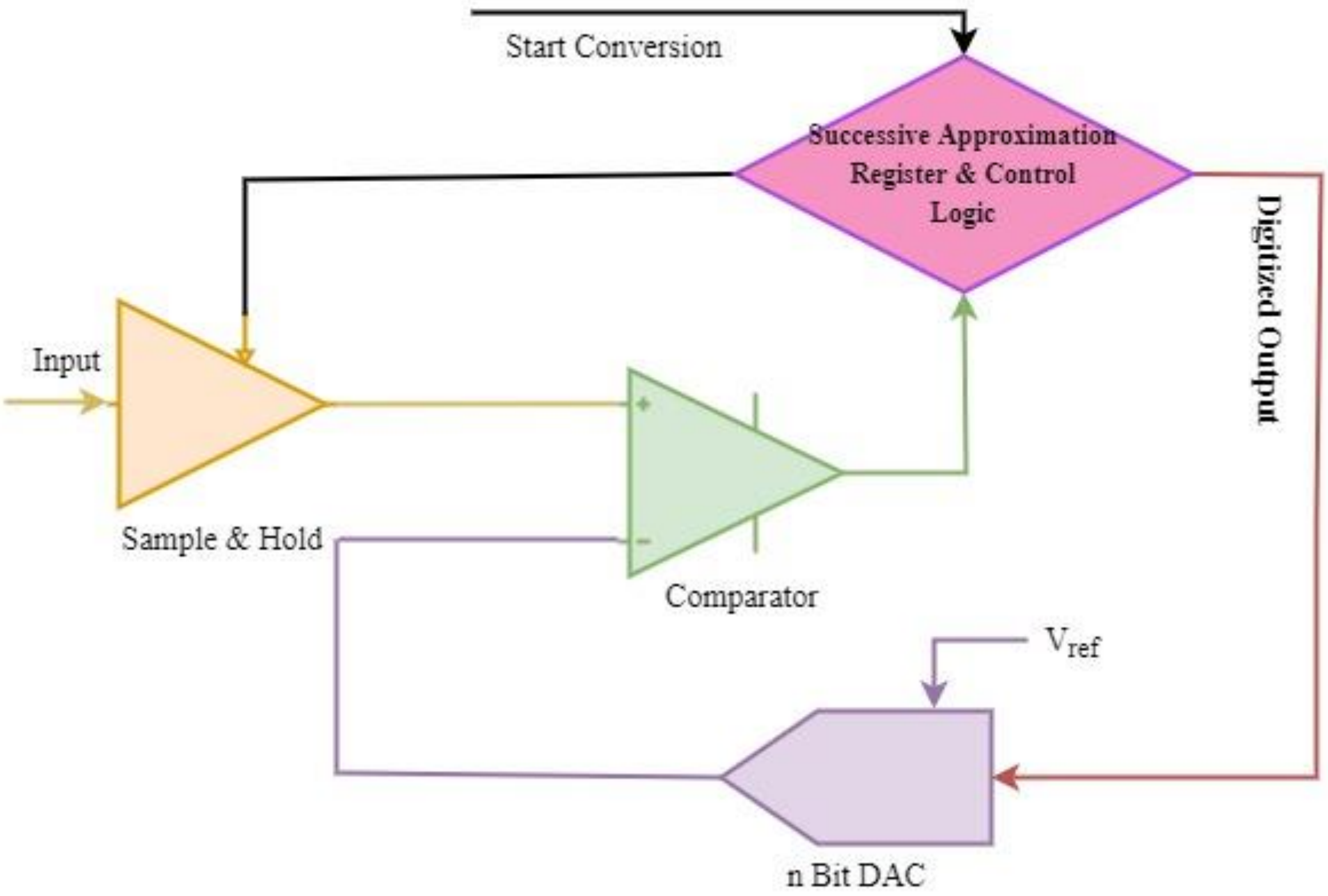

Figure 2: SAR analog to digital converter block

*2.1.2 Sigma Delta ADC (ΣΔ ADC)*

In some applications, such as audio, instrumentation, telecommunication, healthcare, and factory automation & control, the resolution required by the ADC in the signal acquisition chain can reach up to 32 bits. In these cases, the speed of the converter is usually from tens of Hz to a few mega samples per second (MSPs) [14]. ΣΔ converters are best suited for these kinds of applications in terms of resolution and sampling rates. Theoretically, an ΣΔ ADC uses digital filtering, noise-shaping, and oversampling to reduce the quantization noise that exists in the signal bandwidth. The act of sampling the incoming signal more than twice its bandwidth from the Nyquist limit at a rate is known as oversampling. The sampling frequency increases as the noise level in the band of the signal decrease since the overall noise power must remain constant. The signal can be effectively converted by first converting it at a high sampling rate, filtering it with a low-pass filter (LPF), and then decimating the received signal to change the sample rate back to the Nyquist criterion. Figure 3 depicts the structure of the sigma-delta converter. The modulator is made up of an integrator that is driven by the resultant of the input signal and the DAC output. The integrator's output is converted by an ADC before being reconverted into an analog signal by the DAC. The DAC and ADC can both be one-bit converters. A significant advantage of this type of ADC is that the data stream coming from the modulator is filtered by a digital filter. As a result, it is possible to achieve high roll-off while also having more design flexibility in the filter. ΣΔ ADC is the suitable choice for high resolution (larger than 12 bits) at a low-medium sampling frequency (tens of Hz to hundreds of kHz).

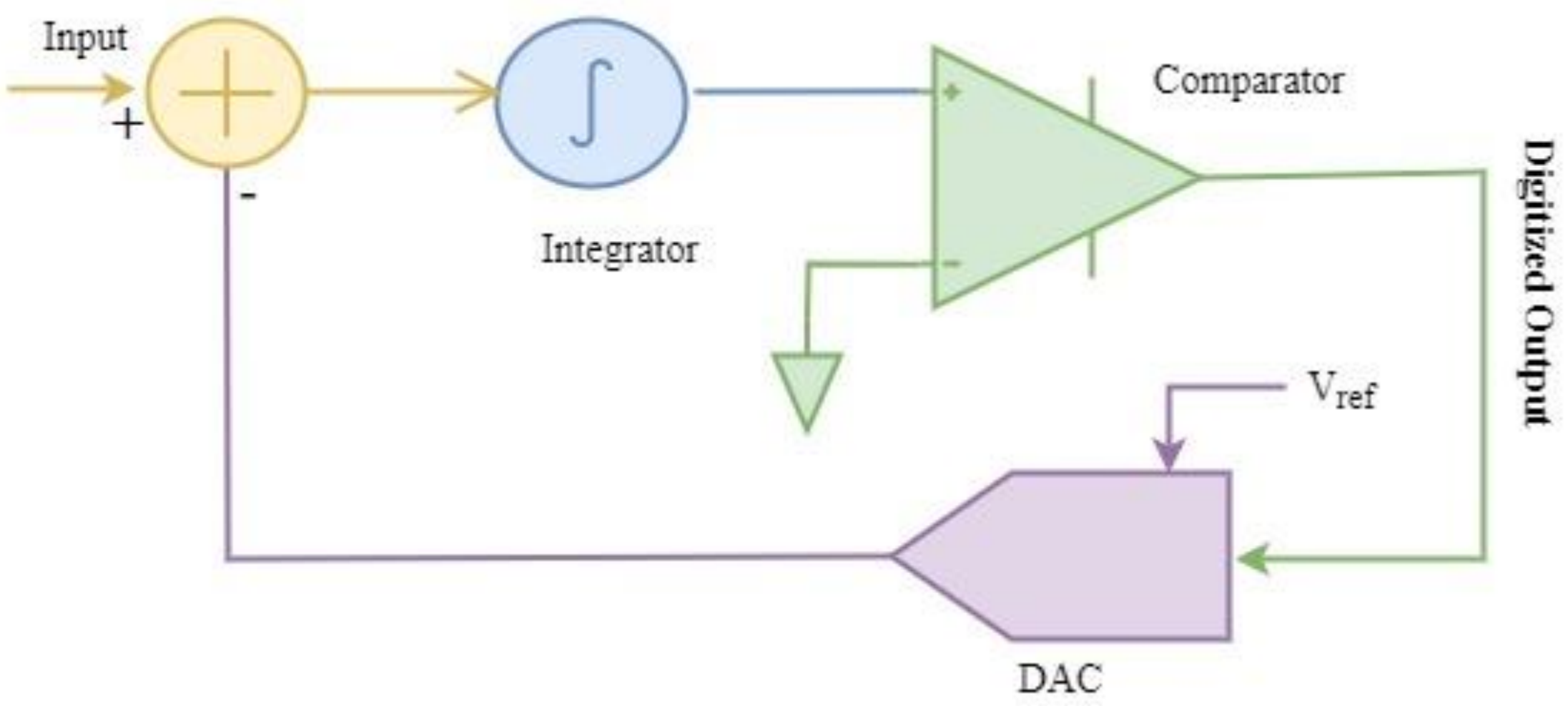

Figure 2: Sigma Delta analog to digital converter block

*2.1.3 Pipelined ADC*

The pipelined A/D converter has been introduced as a best-suited architecture for sampling rates from $10^6$-$10^9$ samples per second. This ADC is working with a higher sampling rate for initial 10-12 bits and later bits are sampled with slower rates. Such resolutions and sample rates cover a broad array of applications, including digital radios, Televisions (HDTV), modems, high-speed Ethernet, biomedical imaging, digital communications, and surveillance systems. However, pipelined ADCs using a hybrid approach have progressed significantly in this decade in terms of data sample rate, precision, high speed of operation, and low energy consumption per conversion. In many applications, the data latency of pipelined ADCs is not a matter of concern [15]. In Fig. 3, the physical world signal $V_{in}$ first goes through a sample & hold block, the flash converter block in stage one quantizes the sampled signal, and the output is then supplied to a digital-to-analog block, where the DAC output is subtracted from the sampled signal. The desired component then gathers this “residue” and feeds it to stage 2. This built-up residue is supplied into the digital error correction logic after the entire bit frame of a given sample has been time-aligned with the help of a register bank using shift registers. A stage can begin to process the next sample it receives from the sampled-and-held output embedded inside the individual stage once it has finished processing the previous sample, deciding the bits, and sending the residue to the following step. The high throughput is a result of the pipelining process.

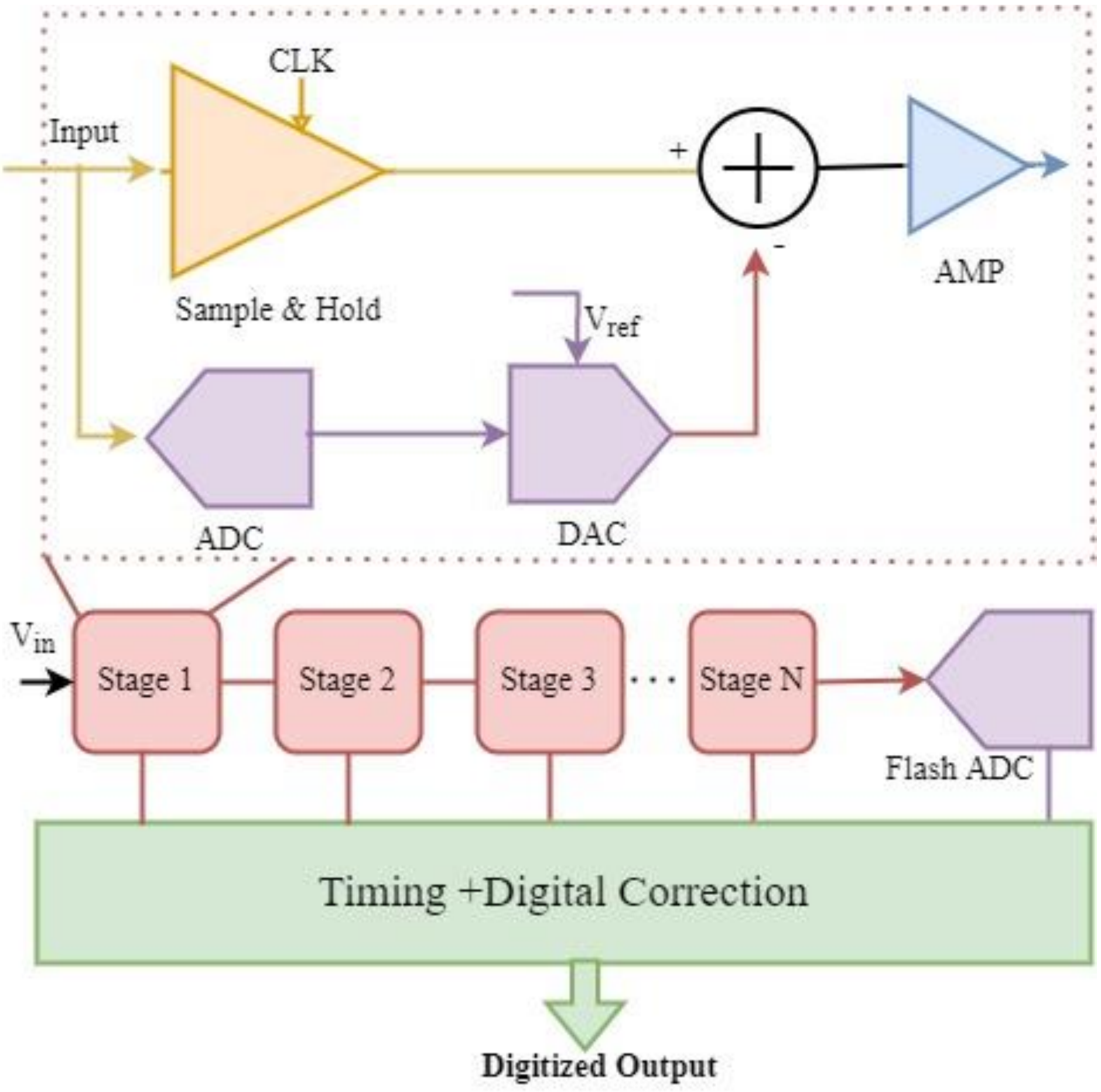

Figure 3: Pipelined analog to digital converter block

*2.1.4 Flash ADCs*

The conversion function is simply carried out by the Flash ADC, as described in fig. 4, by comparing the real-world analog signal with reference values of the resistive network of each quantization interval. The comparison gives logic high for larger non-inverting input and logic low for the high value of inverting reference input. In an n-bit flash architecture, there are $2^n$ regions separating the converter's full-scale range and $2^n$-1 transition points. As a result, $2^n$-1comparators are required to perform the comparisons. All the comparators are activated in parallel and synchronized by a clock signal, thus the output code is generated in one clock cycle. One input of the comparator receives the signal to convert while the other terminal is connected to a reference voltage generated by a resistive divider. The outputs of the comparators form an encoder that can be translated into a digital word. Flash ADCs are mostly employed for high-speed (GHz) converter application like radio, ultra-wide-band, and Wifi since the conversion time is equivalent to a clock period.

However, they are limited by various issues when the resolution becomes larger than 8 bits. As bits increase, the number of comparators also increases exponentially. Area and power consumption are, thus, doubled for each additional bit. Moreover, the resistance value of the unit resistance of the resistive divider is reduced to a very low value. For this reason, the output impedance of the reference voltage driving the divider needs to have a very low value in all the frequencies of operation. In a few words, this architecture is a good choice for high-speed (larger than hundreds of MHz) applications and low resolutions (lower than 8 bits).

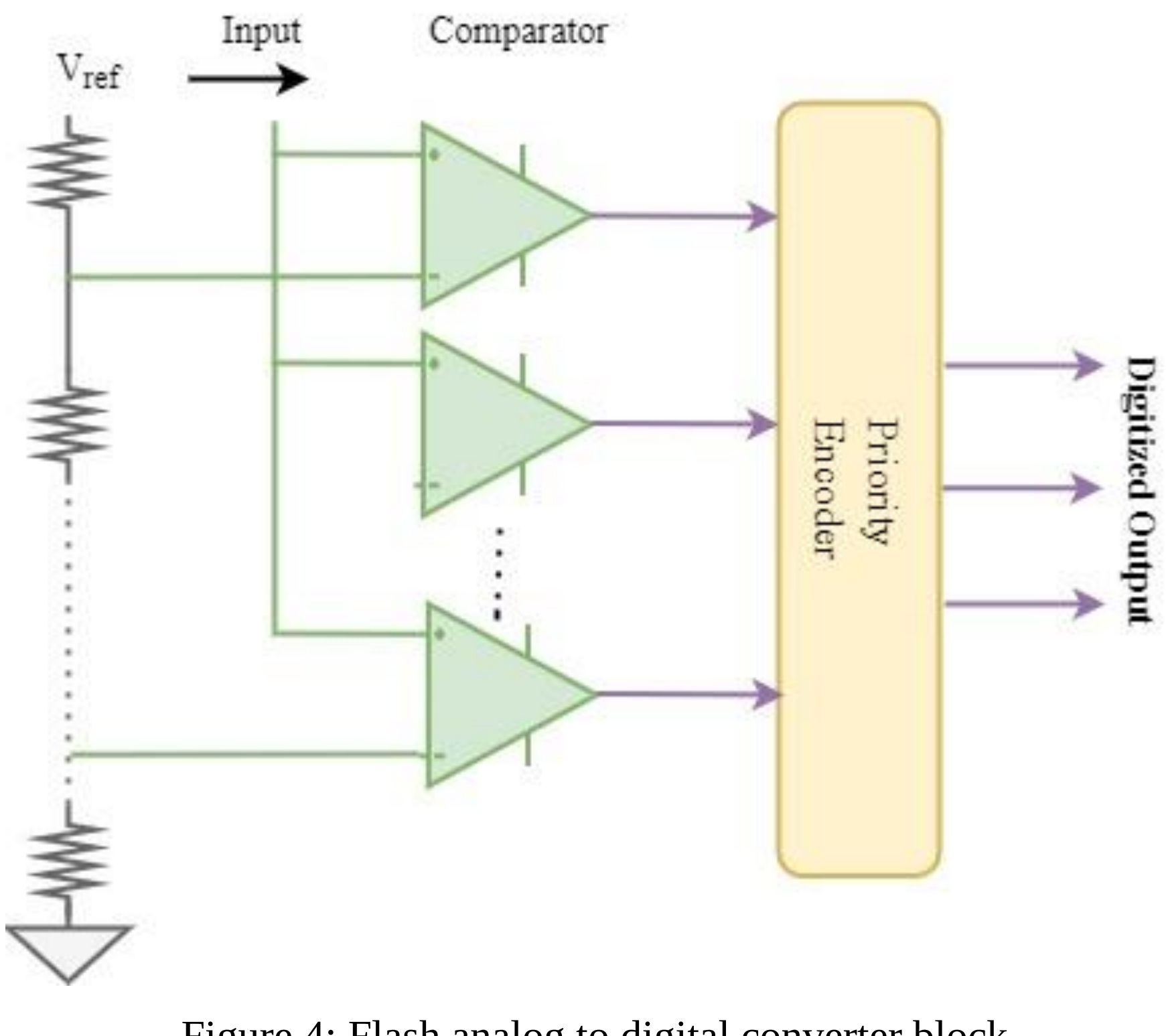

Figure 4: Flash analog to digital converter block

*2.1.5 Digital-to-Analog Converter (DAC)*

A device called a DAC converts a digitized value into an equivalent analog output signal. A digital signal is represented by a digital binary code, which is an arrangement of bits 0 and 1. Various architecture exists in the literature have illustrated in the following table 1. Each architecture has its design and performance benefits.

Table 1: Typical DAC Summary

| DAC Type | Pros | Cons |
|---|---|---|
| Resistor String* | • This network contains $2^N$ resistors in series.<br>• This DAC architecture is simple and guaranteed to be monotonic. | • For high resolution, resistors requirement is high hence increased area.<br>• Due to resistors mismatch linearity errors occurs.<br>• Parasitic capacitance limits the converter's speed since $2^N$ resistors are parallelly connected to achieve higher resolution. |
| Binary Weighted Resistor Ladder* | • Fast conversion due to N resistors in DAC.<br>• Each resistor in this network is a multiple of 2 with each descending bit | • As resistor mismatch occurs in process corner analysis, difficulties with variable resistor values arise for increased resolution. |
| R-2R Ladder* | • A modified version of the binary-weighted resistor ladder with 2N+2 resistors.<br>• Precisely build as only two resistance values R,2R required.<br>• Stray capacitance's negative effects are eliminated by constant node voltage. | • Slower conversion rate<br>• The use of Op-AMP in structure restricts the bandwidth<br>• Occurrence of linearity errors |
| Current Steering* | • Useful in a high-bandwidth requirement<br>• Without Op-AMP performance improve | • No. of current sources leads to a large glitch in the output<br>• High power requirement<br>• Converter’s speed limit by large parasitics |
| Charge Scaling** | • Ease of implementation with capacitors like R-2R, binary-weighted<br>• Performance efficiency in terms of speed, and precision as technology scaling | • Op-Amp based architecture<br>• Capacitors lead to leakages hence accuracy loss occurs after a few milliseconds |
| Oversampling** | • To increase SNR and resolution (ENOB)<br>• The data sample rate is much higher compared to the Nyquist sample rate | • Setup & Hold time issues<br>• Higher power consumption<br>• High Cost |
| Charge Redistribution** | • Lower complexity<br>• Reduced switching energy and area | • Accuracy affected by sampled noise<br>• Nonlinearity due to mismatch |
| Two Capacitor** | • Reduce the effect of mismatch<br>• Noise shaping improves the linearity | • Takes multiple clock cycles to generate a new sample |
| Switched Capacitor** | • Achieve high resolution and high speed | • Higher dynamic energy consumption |

Resistive DAC structure→*, Capacitive DAC Structure→**

### *2.1.6 Summary*

In this chapter, various data converter architectures like a flash, SAR, Sigma delta, and Pipelined structure have been discussed as they are playing a crucial role in major IoT applications. Table 2 summarizes all existing converters with their design trade-offs like sampling speed, resolution, latency, accuracy, conversion time, chip area, power, cost, etc.

Table 2: Typical ADCs Design Tradeoffs Summary

| ADC Architectures | Sample Rate / Speed (Sample/sec.) | Resolution/ Bits | Latency | Accuracy | Conversion Time (No. of Cycles) | Area | Power | Price |
|---|---|---|---|---|---|---|---|---|
| **Flash** | High (1 Gbps-10 Gbps | Low (6-8 Bits) | Low | Low | 1 | High | High | High |
| **Integrating** | Up to 100 | Medium (12-16 Bits) | Low to Moderately high | Mod. High to High | $2^{N+1}$ Dual slope<br>$2^{N}$ Single slope | Low | Low | Low |
| **Counter** | Low (1 to 1K) | Medium (10-12 Bits) | Medium | Moderately high | Depends on amplitude | Low | Moderately high | Low |
| **Pipelined** | Mod. High to High (10 M-5 G) | Mod. High to High (12-18 Bits) | High | Moderately high | $2^{(N/2)-1}$ | High | High | High |
| **SAR** | 10 K to 10 M | Moderately high (8-20 Bits) | Low | Moderately high | Variable | Low | Ultra-low to Low | Low to Moderately high |
| **Sigma Delta** | Low (10 to 1 M) | High (16-32 Bits) | High | High | High | Medium | Low | Low to Moderately high |
| **Interleaving** | Mod. High to High (100 M-5 G) | Mod. High to High (12-18 Bits) | Low | Moderately high | Variable | High | High | High |

## *2.2 Figure-of-Merits*

This section first addresses the various definitions, and requirement and presents state-of-the-art IoT nodes. The selection of the A/D converters for a specific circuit is based on the requirements and application of that design. For purposes of recording, visualizing, and analyzing data, the analog outputs of piezoelectric sensors, motion sensors, temperature and displacement sensors, and more are converted to produce digital data. Important parameters are power, precision, latency, resolution, and speed. The data converter's accuracy can be specified by ENOB which can be calculated from the eq. of SNDR [16].

$$SNDR = 6.02 * ENOB + 1.76[dB] \quad (1)$$

The converter's speed can be illustrated by sampling frequency (fs) and the signal bandwidth (BW). For a Nyquist converter, the sampling frequency is twice the BW. But in the case of Oversampling, BW is much lower than the sampling frequency. In summary, accuracy speed, and power can be represented in terms of figure-of-merits (FOMs) and these FOMs can be calculated with the help of eq.1-3. Figure 5 shows the Walden figure-of-merit (FOMw) and fig. 6 describes Schreier figure-of-merit (FOMs) depiction of more than two decades. These envelopes shown with dotted lines reveal the clear path for designers and show the scope of work.

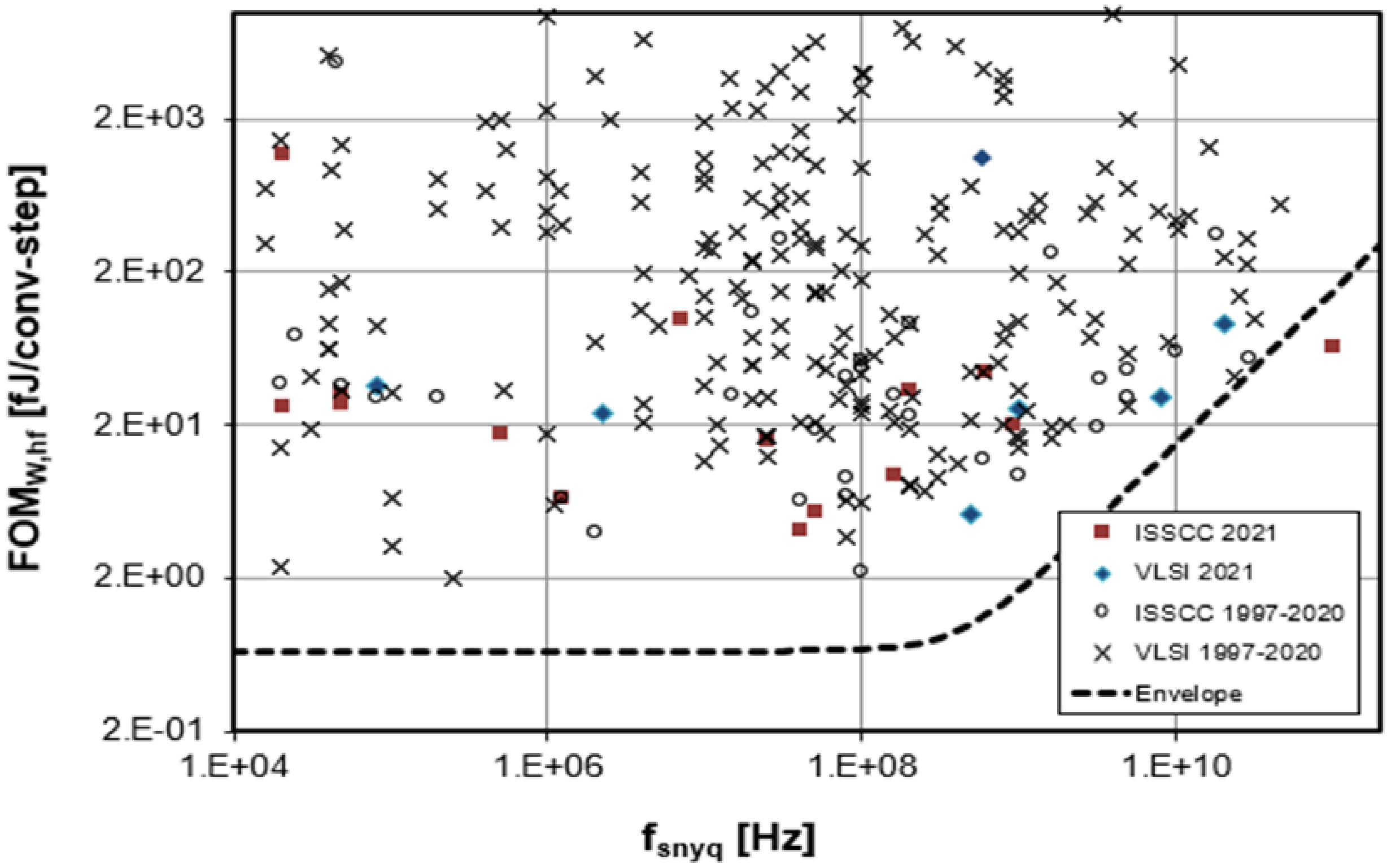

Figure 5: Walden figure-of-merit versus Speed

$$FOMW = \frac{P}{f_{s,Nyquist}.2^{ENOB}}[J / Conversion - step] \quad (2)$$

$$FOMS = SNDR + 10.\log\left(\frac{f_{s,Nyquist}}{2P}\right)[dB] \tag{3}$$

Where P: Power consumed by ADC, ENOB: Effective number of bits (Measure of accuracy), $f_{s,Nyq}$: Clock frequency at which the ADC collects and converts I/P data, SNDR: Signal-to-noise distortion ratio.

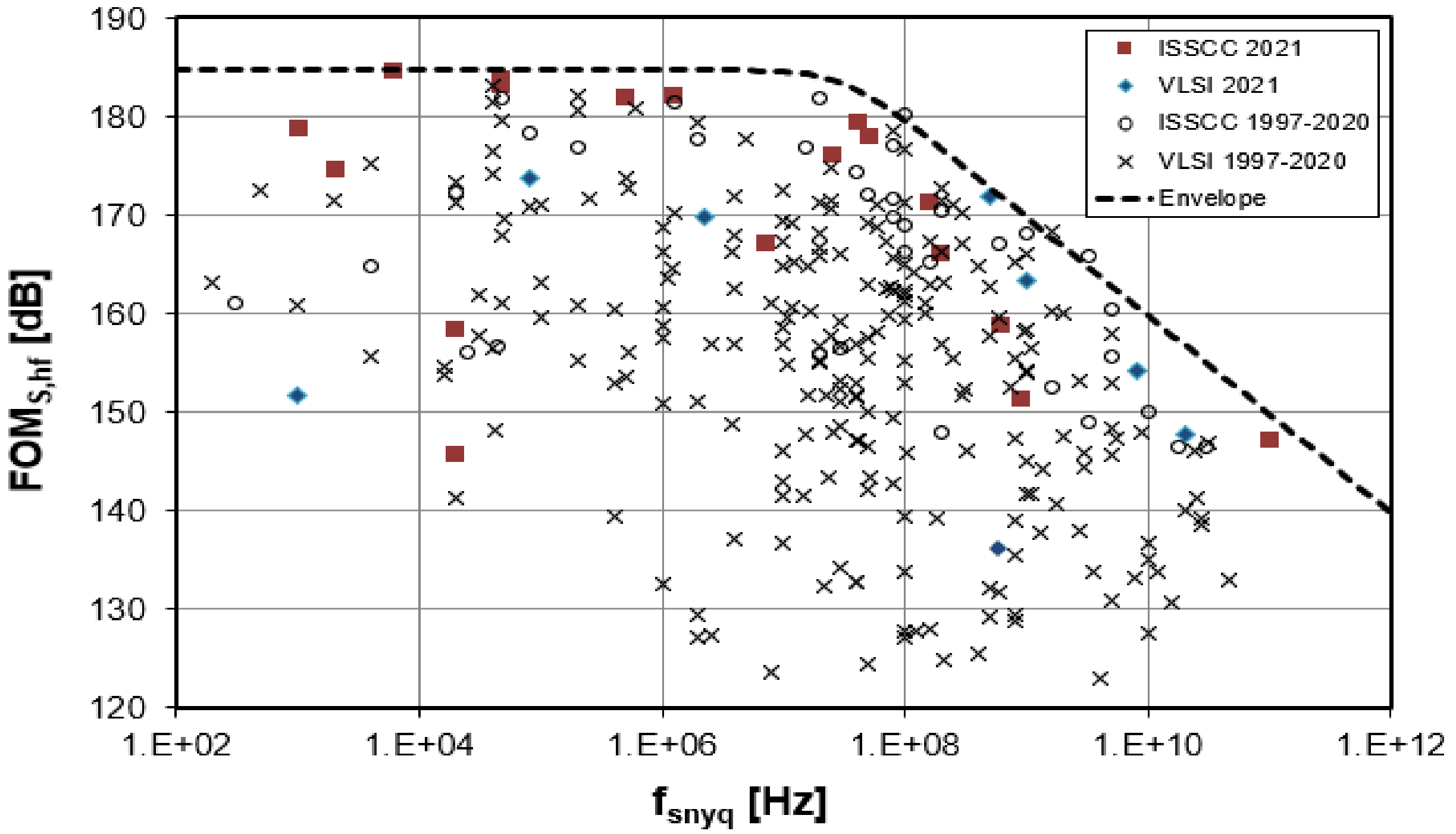

Figure 6: Schreier figure-of-merit versus Speed

*2.3 Present State-of-Art*

Murmann nicely summarises the trends in energy costs for data converters based on available data. [16]. Figure 6 reveals that for low-frequency data converters, the Schreier figure of merit trend increased from 162 dB to 184 dB in the last two decades. Recent research efforts in the field of information converter architectures are depicted in fig. 7 alongside typical Internet of Things (IoT) application areas. This depiction also makes it clear that SAR, which is what the majority of these applications employ, is the best data converter having high power efficiency. In the later section, various analog-to-digital converter designs for potential healthcare applications are presented.

ADC power efficiency improvement over the years is shown in fig. 8. It clearly states that the current research is going on to make the IoT edge node energy efficient. In the overall review, fig. 9 illustrates the energy reduction for converters based on oversampling as well as the Nyquist criteria. Similarly to the design, the converter architecture with accuracy has shown in fig. 10 with the trendline of the present state-of-art.

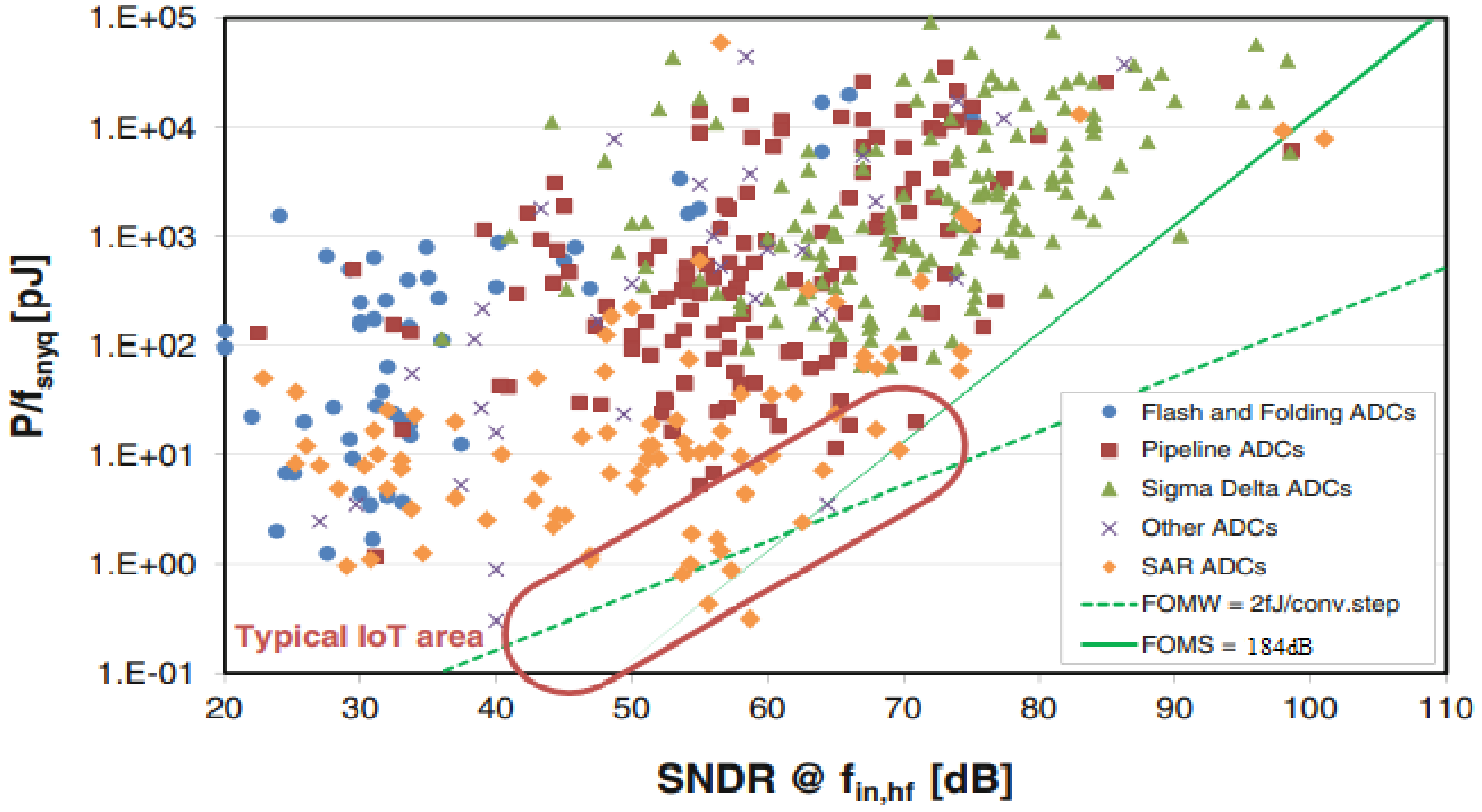

Figure 7: Data Converter power efficiency (Energy) versus SNDR [16]

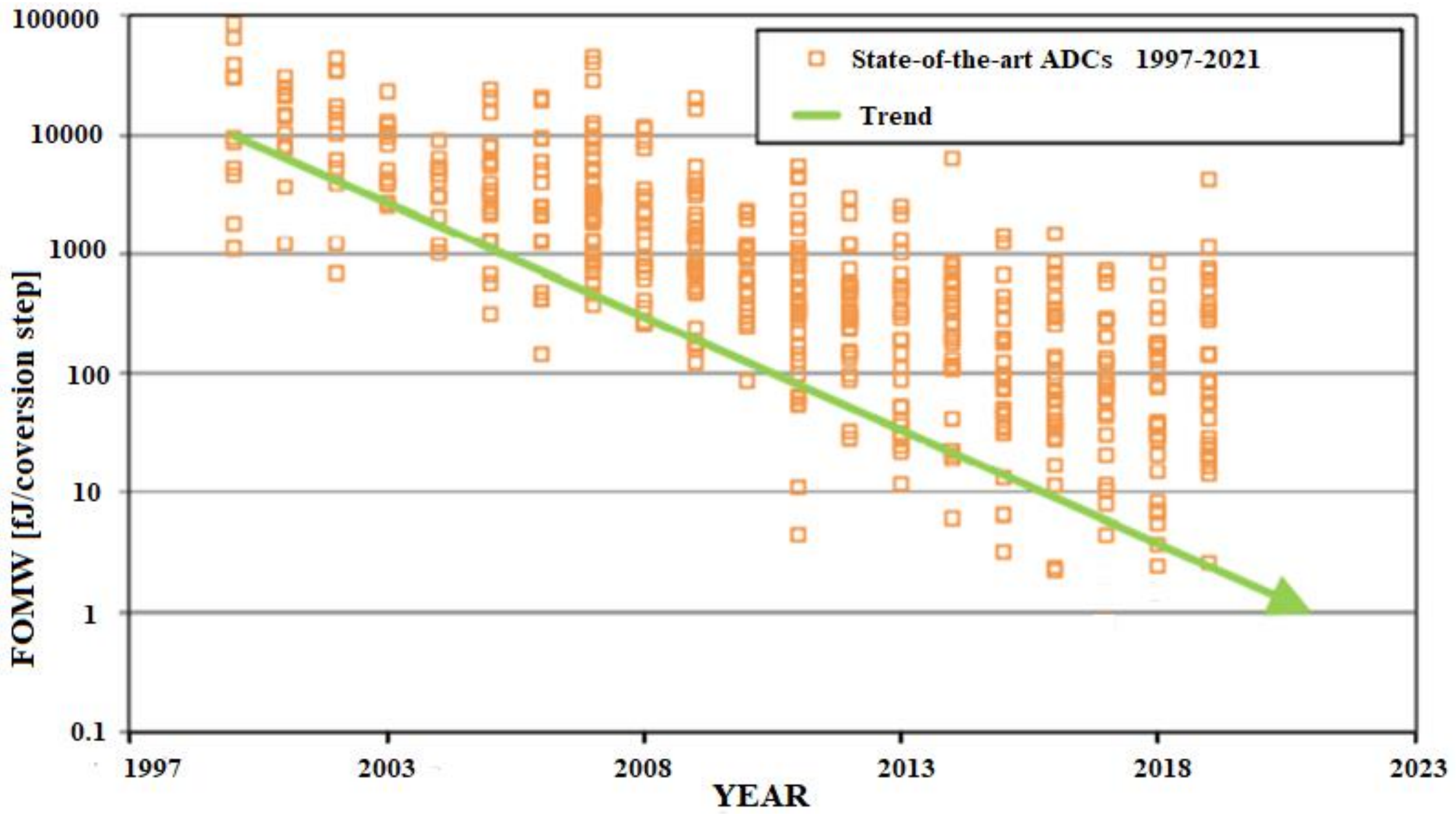

Figure 8: ADC power efficiency improvement over the years [16]

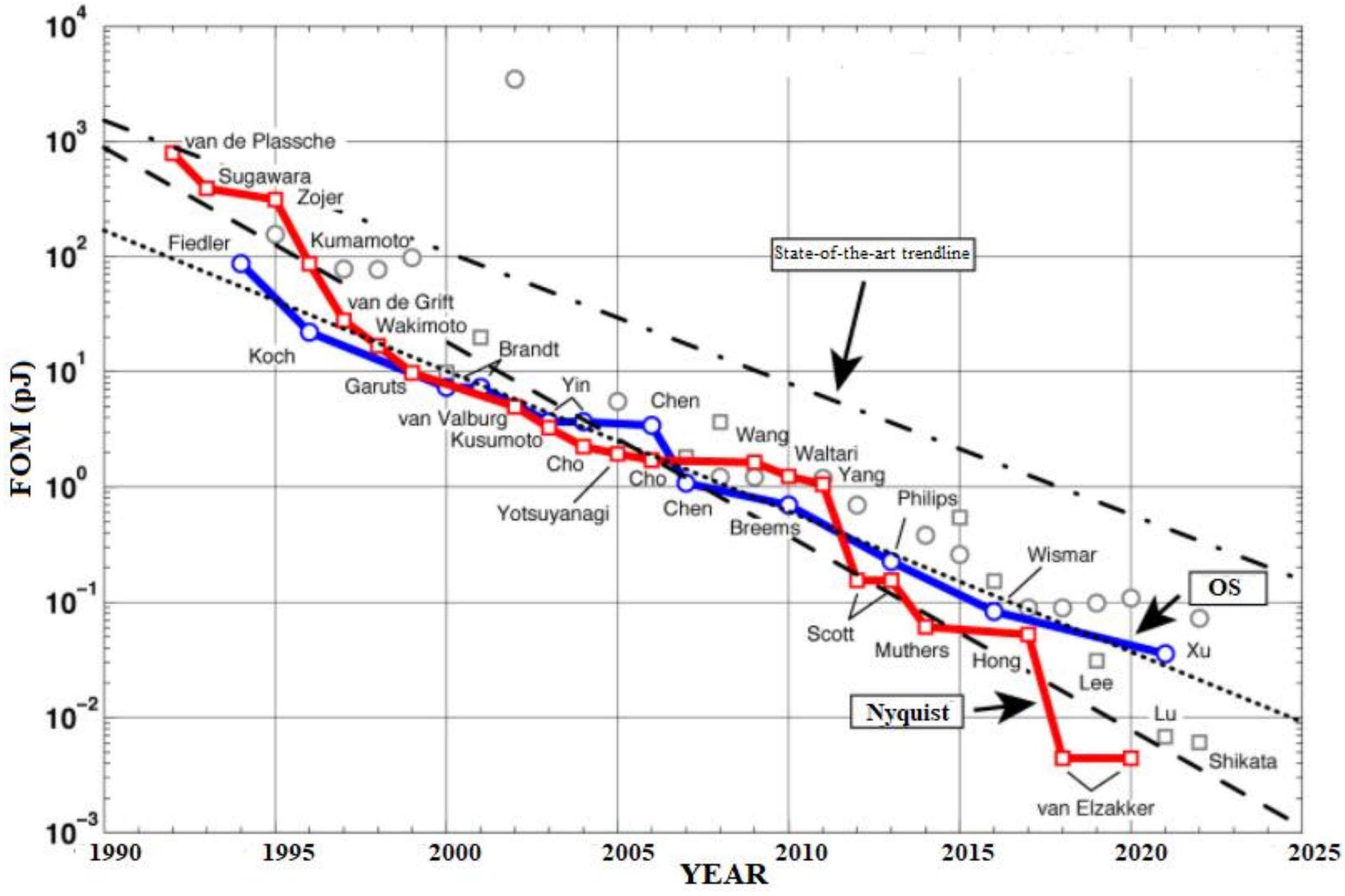

Figure 9: Energy vs Trends

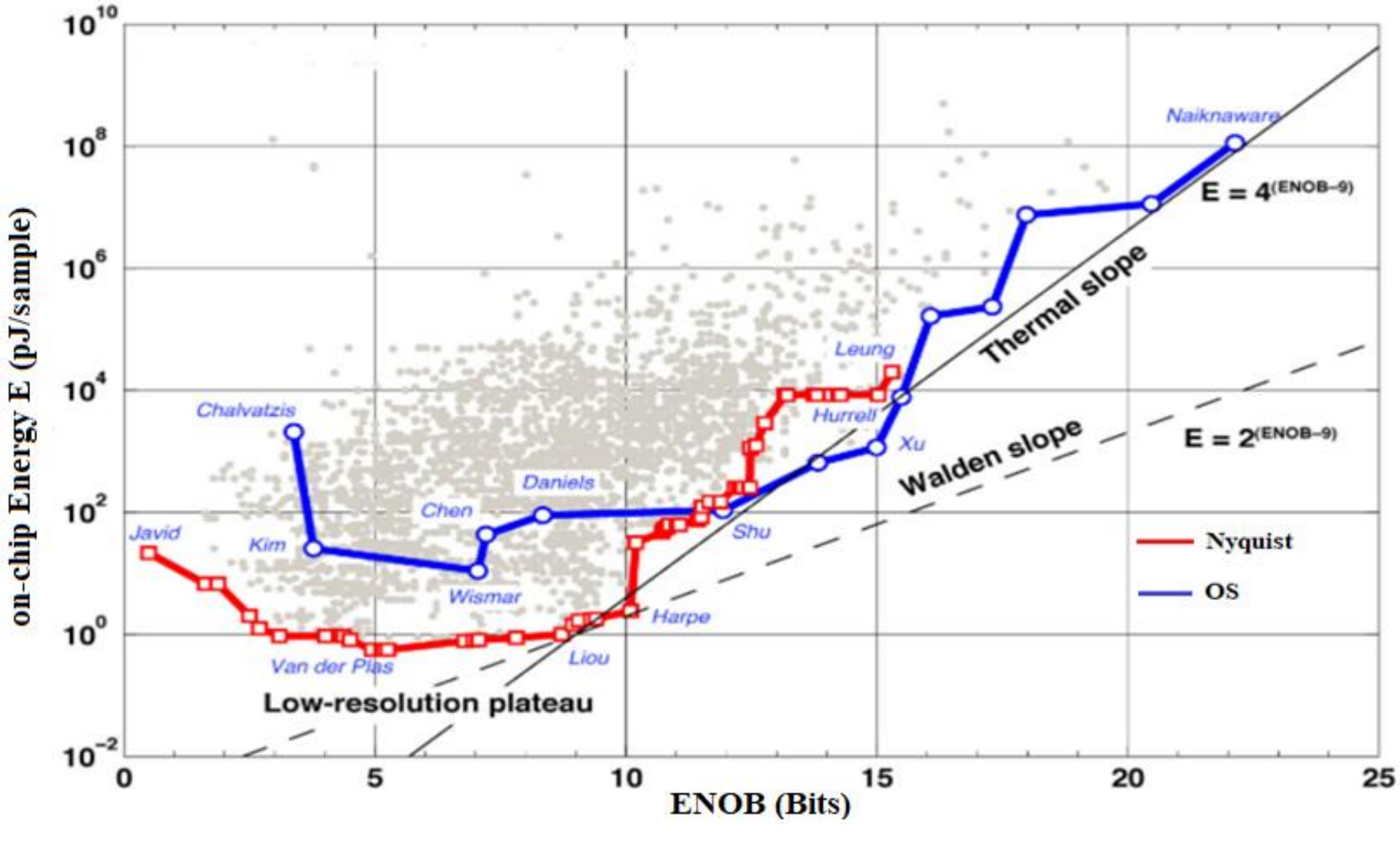

Figure 10: Energy vs Accuracy

# III. Data Converters for IoT

Data converters are found in the sensor interface and wireless receiver. Due to energy consumption limitations, these nodes are optimized for power by compromising accuracy, and speed. There are a variety of applications with varying frequency ranges and resolutions that required different converter architectures as mentioned in table 2. Table 3 shows the numerous biomedical signals from the perspective of the internet of healthcare things (IoHT). IoT-enabled devices have altered the landscape of the healthcare industry's use of remote monitoring, releasing the perspective to keep patients healthy and granting clinicians the right to provide the best care. Patient consultations with doctors are now much simpler and more effective. In addition to reducing hospital costs and lengths of stay, remote health monitoring improves treatment outcomes. Undoubtedly, IoT is revolutionizing the healthcare business by redefining the area occupied by devices and human interaction in the delivery of healthcare solutions. In addition, it has healthcare requirements that assist patients, families, friends, relatives, medical practitioners, hospitals, physicians, and insurers.

## *3.1 Motivation*

IoT is a part of routine life and exists in a variety of applications such as smart grids/houses/cities/retail/wearable, health care monitoring of plants/humans/animals/agronomics, embedded industrial automation, motion/activity monitoring, and environment monitoring. In most scenarios like the Internet of healthcare things (IoHT), these IoT nodes such as wireless frontends and sensor interfaces are inaccessible, and operating renewal is not advisable. Table 3 shows the suitable data converter architecture according to the IoT healthcare application. Again we see the SAR ADC is the most suitable guy as it is covering maximum biomedical application areas.

Table 3: IoT for Healthcare

| **Signal** | **Frequency Range** | **Data Converters** | **Application** | **Remarks** |
|---|---|---|---|---|
| Electrogastrogram (EGG) [17] | DC-1Hz | Sigma Delta | Motility Disorder Monitoring | -Requires 32-bit resolution<br>-Requires low sampling rate |
| Respiratory rate [17] | 0.1-10Hz | Sigma Delta | Heart Monitoring | -Requires 10-bit resolution |
| Electroretinogram (ERG) [17] | DC-50Hz | SAR, Sigma Delta | Eye Monitoring | -Sampling rate 1KHz |
| Blood Pressure (BP) [17] | DC-60Hz | SAR | Blood Pressure Monitoring | -Requires 8-bit resolution |
| Electrooculogram (EOG) [20] | DC-100Hz | SAR, Sigma Delta | Eye Monitoring | -Requires 16–24-bit resolution |
| Electroencephalogram (EEG) [17] | DC-150Hz | Sigma Delta | Brain Monitoring | -Requires 12-bit resolution |
| Electrocardiogram (ECG) [17] | 0.01-250Hz | Sigma Delta | Heart Monitoring | -Requires 16-bit resolution |
| Electroneurogram (ENG) [17] | 250Hz-6KHz | SAR | Brain Monitoring | -Requires 10-bit resolution |
| Electromyogram (EMG) [17] | 20Hz-1KHz | SAR | Muscles Monitoring | -Requires 12-bit resolution |
| Phonocardiogram (PCG) [17] | 20Hz-20KHz | Sigma Delta | Heart Monitoring | -Requires more than 16-bit resolution |
| Photoplethysmogram (PPG) [21] | 0.5-5Hz | Sigma Delta | Heart Rate Monitoring | -Requires up to the 22-bit resolution |
| Positron Emission Tomography (PET) [18] | >40MHz | Fully differential SAR | Medical Imaging | -Requires 10–12-bit resolution<br>- Low noise as well as power to enhance the dynamic range & reduce heat dissipation |
| Magnetic Resonance Imaging (MRI) [19] | 12.8MHz -298.2 MHz | low power multichannel pipeline | Medical Imaging | -Requires 16-bit resolution<br>- Oversampling the MR signal to achieve enhanced image<br>-increases SNR as well as eliminates aliasing artifacts |
| Ultrasonography [18] | 1MHz-18MHz | SAR, Pipelined | Medical Imaging | - Requires high ENOB, high speed, and low THD (total harmonic distortion) |
| Computed Tomography (CT) [19] | 0.4-20Hz | Sigma Delta | Medical Imaging | -Requires 24-bit resolution |
| Digital Radiography (X-Ray) [19] | 60MHz-100MHz | SAR, Pipelined | Medical Imaging | -Requires 14–18-bit resolution<br>-SNR level 70dB-100dB<br>- Multiple ADCs with this sampling rate |

This forces the IoT architecture to have a long battery life with less maintenance. Energy-efficient IoT architecture blocks enable the long autonomous operation and reduce the size of the battery or harvester used to give energy to the IoT edge node, making their incorporation into large-scale healthcare monitoring systems easier. Although the signals sensed in the real world are required to be transformed to the domain with digital information. Thus, Analog-to-Digital converters (ADCs) and Digital-to-Analog converters (DACs) form a requisite component in IoT devices.

### *3.2 Growth & Marketplace*

Human lives are improving with the widespread use of cutting-edge digital technology like the Internet of Things (IoT). Recently, the pandemic has shown the demand for more digitally advanced IoT-based devices. International Data Corporation (IDC) forecasts that by 2025, there will be approximately 42 billion of these devices in use, capable of producing around 80 ZB (zettabytes) of data. The market segmentation of data converters is based on type, industry vertical, and geography. ADCs and DACs are the two broad categories of converters available in the market. The market is divided into aviation and security, industrial automotive, communications, electronics, healthcare instrumentations, etc. by industry vertical sector. The market for data converters is geographically dominated by North America in 2020, with a 37.5 percent market share. Because of the rise of the telecommunications, consumer electronics, and automotive industries, North America currently dominates the market. As a result, the Asia-Pacific area has a lot of room to grow in this sector.

From 2021 to 2026, the data converter market is expected to achieve 5.1 billion dollars, increasing at a CAGR (Compound Annual Growth Rate) of more than 5 percent. The market has grown due to the increasing usage of modern automatic acquisition systems and the demand for high-precision images in scientific and healthcare applications. According to IDC (International Data Corporation), the global IoT industry revenue would be around US$1.1 trillion between 2021 and 2025. The number of IoT connections is estimated to rise at a 17 percent CAGR from 7 billion in 2017 to 25 billion in 2025. Cars will be the fastest-growing application (30 percent of the M2M share by 2023) situation in recent IoT referred to M2M (mobile to mobile) connections category connected home applications (50 percent CAGR of the M2M share by 2023).

### *3.3 ADC/DAC requirement for IoT*

Sensors and industrial applications as addressed in the section, power requirement in IoT devices, IoT networks have made rapid progress in terms of system performance over the last two decades. As a result, Table 4 shows the power consumption of several healthcare IoT devices and applications that are currently available.

Table 4: Power Consumption requirements for healthcare IoT (IoHT) [22]

| **Application** | **Performance** | | |
|---|---|---|---|
| | Power Consumption | ADC/DAC | Energy Source |
| Heart Monitoring | <10μW | 0.8-1.2KSPS, Sigma Delta ADC | Battery with more than 10 years of lifespan |
| Body area Monitoring | <140μW | 1KSPS, 12-bit ADC | Battery |
| Hearing Aid | 100-2k μW | 10-15 KSPS, 10-12 bit ADC | Rechargeable battery with one week lifetime |
| Eye Monitoring | 250mW | 10KSPS, 4bit DAC | Inductive power |
| Brain Monitoring | 1-10mW | 100KSPS, 8-10 bit ADC | Inductive power |

To achieve extremely energy-efficient operations, data converters with a resolution of lesser than 10 bits are extensively employed in AFE circuits. Greater resolution (≥10 bits) and ultra-low-power consumption are two conditions that must be met by these converters before they can be employed in high-end battery-powered devices. Higher sampling speed, Lower power consumption, lower supply voltage, and higher resolution are becoming more important to data converter designs as the channel length of the transistor shrinks (Technology scaling).

### *3.3 Challenges and future directions*

Research into the design and development of various data converters is still ongoing, according to a literature review [23-38]. Ultra-low energy data converter designs for the best speed and accuracy in the low to the moderately high-frequency range are some research gaps that need to be filled. Most data converter implementations concentrate on improving key metrics. Precision is one of the fundamental elements requisite by plenty of IoT applications to predict and forecast the desired value. Real-time application requirements are incompatible with the capabilities of any converter circuit. As a result, a high-resolution figure of merit must be established for such applications.

Rapid advancements in IoT technology are forcing designers to create new designs. Signal-to-noise-distortion ratio, sampling rate, resolution, power consumption, ENOB, and spurious-free dynamic range (SFDR) are all important design characteristics. Although data converters are used in many different industries, integrating them into system-on-chips (SoCs) and field-programmable gate arrays (FPGAs) is a challenging task. It takes a lot of skill to implement FPGAs and SoCs in smart IoT devices. These factors attract professionals who are likely to contribute.

Infrastructure, agriculture, utilities, home automation, healthcare, automotive, industrial, and other applications are all possible with IoT devices. The entire spectrum of potential applications has not been considered. As more applications are addressed, the market for IoT devices will explode. There will be a lot of fresh ideas in this field, and a lot of them will originate from businesses without a history or track record in creating high-tech tools and machinery.

A single system on a chip (SoC) will be used to implement many of these new IoT innovations, which will present challenges in terms of providing the highest level of integration and chip area reduction. There may be room for customization and differentiation in high-performance analog and mixed-signal blocks. To benefit from both power and chip area savings, the majority of IoT SoC designs are implemented with technology scaling. However, due to transistor mismatch and leakage, there are significant challenges in this domain. This chapter illustrated various converter architectures for healthcare IoT, we can expect these topologies to continue pushing energy efficiency trends.